# The Key Elements of Viral Advertising. From Motivation to Emotion in the Most Shared Videos

Dafonte-Gómez, Alberto

## Abstract

From its origins in the mid '90, the application of the concept of virality to commercial communication has represented an opportunity for brands to cross the traditional barriers of the audience concerning advertising and turn it into active communicator of brand messages. Viral marketing has been based, since then, on two basic principles: offer free and engaging content that masks its commercial purpose to the individual and use of using a peer-to-peer dissemination system. The transformation of the passive spectator into an active user who broadcasts advertising messages promoted by sponsors, and who responds to needs and motivations of individuals and content features which has been described by previous research in this field, mainly through quantitative methods based on user perceptions. This paper focusses on those elements detected in its previous research as promoters of the sharing action in the 25 most-shared viral video ads between 2006 and 2013 using content analysis. The results obtained show the most common features in these videos and the prominent presence of surprise and joy as dominant emotions in the most successful viral videos.

[English PDF file](#)

[Spanish PDF file](#)

## 1. Introduction and State of the Art. Viral Video Advertising and the Role of an Active Spectator

Viral video advertising may be the most popular manifestation of viral marketing phenomena. The concept of virality, introduced in the field of media theory by Rushkoff (1994)[1], has been rapidly adapted to marketing under the basic principles described by Rayport in 1996, in his now acclaimed seminal article for this new approach[2]: disguising the content's commercial aim and making the users themselves circulate it through their contact networks.

Although «online» videos did exist before this breakthrough, the growth and consolidation of this platform initially targeting the dissemination of clips generated by users («User-Generated Content» or UGC), enlarged audiovisual content distribution channels parallel to traditional media by using the users' networks and small subscriber communities around the channels on this platform. The popularisation of videos was no longer a response to a situation of simultaneous consumption by mass audiences, but to a distribution structure through networks of users who, in an asynchronous way –although generally concentrated in time– share the content amongst their community of contacts. This is the breeding ground for the growth and development of viral videos.

What then is a viral video ad? First, it is a video produced by a brand with a direct or indirect commercial goal. This term is often generically used to refer to videos that have reached a high number of views, but this can be achieved through different means, such as paid content promotion, recommendations of similar videos on YouTube or the amplifying effect of TV broadcasting. If we are to be true to one of the main principles of viral marketing, viral videos must –necessarily– be shared by many individual users. What makes the difference in viral audiovisual content is, beyond the number of views –an important parameter, no doubt– the number of views achieved through mass dissemination by users who share a video across their contact networks of any kind (Porter & Golan, 2006: 29; Eckler & Bolls, 2011: 2). Despite the fact that a successful viral video may reach a large number of views, virals are so designed for their sharing, not for their views. In order to enhance content sharing, viral videos are stored and disseminated through a network. However, at present, viral video ads in particular may be clips that have been broadcast on television, either before or simultaneously. Lastly, another typical feature of this kind of video is that the large number of views implicit in multiple sharing is achieved in a brief period of time – after the initial peak, the growth in hits dramatically decreases (Broxton, Interian, Vaver & Wattenhofer, 2010).

While the number of views achieved is the result of the passive act of watching –an anonymous act that does not mean any personal implication–, the act of sharing means that there is a symbolic link between the content shared, the personality of the user sharing it, and the perception of the community it is shared with. It is therefore natural to reflect upon the motivations that make a simple spectator become a tool for the dissemination of a message, subjecting their likes, preferences and even convictions to the scrutiny of their community through the action of forwarding the content. In this sense, we have to state the obvious – viral marketing does not create this social habit, but just makes the best of a pre-existing audience behaviour (Aguado & García, 2009).

Word-of-mouth or WOM communication has attracted increasing attention in the academic field of marketing and advertising since the 1950s. According to De Bruyn and Lilien (2008: 152) the research on this topic that has developed since then revolves around three main axes: research on the reasons why a consumer disseminates, in a proactive fashion, their consumption experience of a series of products or brands; research on the situations in which consumers trust WOM more than other sources of information before they purchase something; and research on the reasons why the information supplied by some people can have more influence on recipients.

As years have gone by and new media and communicative phenomena have appeared, research on the users' motivations to share their experience with a brand through word-of-mouth have given way to studies that, without forgetting their origins, focus on the reasons to share content of any kind through email or social networks, and later, on ad content, or more specifically on viral video ads.

Sundaram, Mitra and Webster (1998), writing about WOM, highlight that we are more inclined to share information on products that we feel are useful for our community, and that we also preferentially share information about those products or brands we are proud to use or about products that we think that define our personality. This thesis is further supported by Chung and Darke (2006), who claim that despite the numerous and varied products that we consume, we only share user experiences about those that we think strengthen the personal image that we want to project.

When we share our experience and interest in products and services that we use, we are taking consumption acts that are, in many cases, private in themselves, into the public sphere, so that the decision to make them public is part of the image that we project towards others.

Ho and Dempsey (2010: 1001) developed a study on the motivations of users when sharing on-line content based on the theories of Schutz (1966) on interpersonal behaviour («Fundamental Interpersonal Relations Orientation»: FIRO) that claim that human beings communicate and relate to satisfy personal needs according to three axes: inclusion, affection and control. The motivations discussed by Schutz are further developed by Ho and Dempsey with the hypothesis that users who consume more online content, and those with higher curiosity levels and a willingness to discover may be more likely to share. The outcomes show that, of the two axes of analysis proposed, only the need for inclusion and affection have a direct influence on the act of sharing content through a network, and it is the need to differentiate oneself from the group –the need to claim that one is distinct and «unique»–, rather than the idea of belonging to the group that triggers the action of sharing content on the web.

However, beyond the motivations related to the construction and projection of the individual's identity, the struggle for acceptance by their surroundings or the contribution to the knowledge of the community, there are reasons that have to do with the content itself and with the perception that the user has of it.

Research such as that by Huang, Chen and Wang (2012: 12) highlights that the quality of content is a determining factor in the decision to share it, while the expected response by the recipient (conceptualised as «empathy» by the authors) has an indirect influence on the hopes for inclusion, control or affection by the user sharing it. From a different perspective on the same aspect, Eckler and Bolls (2011) focus their research on the emotional content of viral messages, and the response they have to generate to become one, although they consider that this is not the only factor explaining the reaction of users, and they argue that generating emotions is a necessary requirement for a video to be shared. They also argue that, as viral videos are thought to be more provocative than conventional TV ads, an analysis of the emotional tone of viral videos and their effect on the attitude of users towards the ad, the brand and also their intention to share it is much needed. Eckler and Bolls (2011: 8) conclude that the emotional tone of viral video ads is directly related to attitudes and intentions, and they highlight that ads that are perceived as pleasurable are more effective in generating positive attitudes towards the brand and interest in sharing than those that are perceived as disagreeable or coercive. This further suggests, according to the authors, that the taste for provocative or controversial content that ad producers attribute to their audiences might not match what the users want.

The idea that viral ads can afford to be –or that they even should be– more dramatic than ads for television is not new, and researchers such as Porter and Golan (2002: 31) already highlighted in 2002 that, although the emotional aspect linked to success of ad content was generally accepted, the success of viral video ads depended more on the excitement they could create through «provocative» and «crude» content, which would be most likely shared by users. The study by Porter and Golan concludes that sex-, violence and nudity-related content – generally enveloped by an air of comedy– gets better results in terms of dissemination by users.

Other authors such as Teixeira (2012) in more recent articles argue that it is essentially content based on happiness and surprise that keeps the spectator's attention, while the decision to share it or not is more related to the individual's personality, highlighting extroversion and/or egocentrism as the traces of users more prone to share content.

Dobele, Lindgreen, Beverland, Vanhamme and van-Wijk (2007) also consider surprise as the fundamental emotion that a viral video must generate to be shared. These authors' research focuses on identifying the perception by spectators of the six primary emotions previously described by Ekman (1972) (surprise, fear, sadness, happiness, disgust and rage) in successful campaigns. Their conclusions emphasize the importance of surprise for the dissemination of viral content, but also that it is usually accompanied by some of the other five emotions, and that the combination of surprise with happiness or disgust (with a humorous outcome) increases the intention of sharing the content by spectators (Dobele et al. 2007: 295-301). However, the authors themselves indicate that the presence of emotions is not enough to force the content to be shared, and that a campaign that «capture[s] the recipients' imagination in a unique or unforgettable way» becomes a must.

Looking at the state of the art, it is obvious that the decision to share a viral video is caused, on the one hand by motivations that have to do with the psychological or emotional needs of the user potentially sharing the clip, and on the other, with the motivations related to the viral video itself. The decision to share a viral video ad stems from the meeting of both these spheres in the individual.

In the face of the conclusions presented by this kind of research –generally of a quantitative nature– based on the perceptions of users about themselves and on the intention of sharing content, one could ask to what extent highly successful viral video ads –that is to say, those that have already been massively shared in an effective way– comply with the features labelled as relevant by previous studies based on the perception or behavioural prediction of users when they face viral content.

The objective of this paper is, first, to analyse highly successful viral video ads and find their common features and, second, to confirm that, indeed, the success of such viral video ads includes the elements highlighted as relevant by previous research on users. If we cross-check the user perspective with the analysis of content we will ascertain the degree of adequacy existing between what the individuals claim that leads them to share a video, and the features of the most successful viral video ads.

## 2. Methodology

The methodology proposed is based on content analysis of a sample of 25 viral video ads of proven success, looking at the elements identified as triggers for the act of sharing them in previous research.

The selection of viral videos was performed –considering the ideas put forward in the introduction– by looking at the number of times a viral video ad was shared and not at the number of views. As we have already explained, we consider that the act of sharing signifies a viewer of a video deeper involvement with the content in comparison to the mere act of watching it. On the other hand, sharing by users is an essential part of the concept of virality.

In order to obtain data related to the number of times a video ad disseminated through the web was shared, we have taking as reference the public ranking created by Unruly Media (n.d.) –a company specialising in marketing for viral videos– in cooperation with Mashable, which is also followed by publications such as «Adweek», «The Guardian» or «The Washington Post», and institutions such as the «Internet Advertising Bureau» (UK), amongst others. We have selected the first 25 video ads in Unruly Media's ranking according to the number of times they were globally shared, from the beginning of their operations (2006) until the present. We have selected the sample for the longest period of time that the platform allows, as we think that, in this way, the results obtained are more consistent than if we focused on shorter and more recent periods, such as the latest week or month in the ranking. We also considered that a longer time-frame would lead to fewer less variations in the list, which is subjected to fluctuation depending on user shares. The data for this study were last updated on 23/11/2013. The list of videos that have been analysed are shown in table 1.

The following data was registered for each item: title, sponsor, agency, year, duration, number of shares, number of views, TV broadcasting, target market, use of celebrities, humour, eroticism, violence, presence (not excluding) of the following emotions: surprise, fear, sadness, happiness, disgust and rage and, lastly, final emotional tone conveyed (agreeable/disagreeable).

Once the analysis was started, we perceived that the item «surprise» could be subjected to a more precise delineation depending on the discursive element used to cause it. Thus, four different types of «surprises» were described in the units analysed: «real stunt» (surprise caused by dangerous scenes, generally of a sporty nature, with stunt actors or experts as their main characters, such as those by DC Shoes, Red Bull or Volvo), «fictional stunt» (surprise caused by activities impossible to perform by the person doing them, usually with the aid of digital effects, for example the clips by Evian or Geico), «surprise event» (surprise caused by a thrilling action of «street marketing» developed in public spaces and recorded with a hidden camera, such as those by TNT Benelux or MGM), and «narrative surprise» (this refers to the narrative turn used in many works of fiction to achieve an unexpected ending, such as those by Volkswagen or Budweiser in 2013).

| | Table 1. Most widely shared viral video ads between 2006 and 2013 | | | | |
|---|---|---|---|---|---|
| | Title | Sponsor | Year | Shares | Views |
| 01 | The Force | Volkswagen | 2011 | 5577,99 | 66,979,031 |
| 02 | A Dramatic Surprise on a Quiet Square | TNT Benelux | 2012 | 4,707,252 | 51,156,565 |
| 03 | Dove Real Beauty Sketches | Dove | 2013 | 4,273,590 | 128,006,741 |
| 04 | Dumb Ways to Die | Melbourne Metro Trains | 2012 | 4,246,095 | 68,084,885 |
| 05 | Hump Day Camel Commercial | Geico | 2013 | 4,051,911 | 18,505,360 |
| 06 | Baby&me | Evian | 2013 | 3,353,756 | 66,215,113 |
| 07 | 9/11 | Budweiser | 2002/08* | 3,345,654 | 14,418,895 |
| 08 | DC Shoes: Ken Block's Gymkhana Five: Ultimate Urban Playground; San Francisco | DC Shoes | 2012 | 3,294,596 | 55,002,809 |
| 09 | DC Shoes: Ken Block's Gymkhana Three, Part 2; Ultimate Playground; l'Autodrome, France | DC Shoes | 2010 | 3,182,060 | 54,165,642 |
| 10 | Evian Roller Babies | Evian | 2009 | 3,173,096 | 118,177,742 |
| 11 | Ship my Pants | Kmart | 2013 | 3,078,971 | 29,972,830 |
| 12 | The Hottest @Abercrombie & Fitch Guys, «Call Me Maybe» by Carly Rae Jepsen | Abercrombie & Fitch | 2012 | 2,918,863 | 20,806,532 |
| 13 | Yalın - Keyfi Yolunda, Alkı Sonunda | Cornetto | 2013 | 2,911,402 | 26,401,796 |
| 14 | DC Shoes: Ken Block's Gymkhana Four; The Hollywood Megamercial | DC Shoes | 2011 | 2,853,791 | 25,149,752 |
| 15 | The Clydesdales Brotherhood | Budweiser | 2013 | 2,719,978 | 15,417,557 |
| 16 | Pepsi MAX & Jeff Gordon: «Test Drive» | Pepsi | 2013 | 2,696,314 | 45,299,800 |
| 17 | The Epic Split feat. Van Damme | Volvo Trucks | 2013 | 2,518,742 | 49,343,772 |
| 18 | 2 year old dancing the jive | Studie43 | 2012 | 2,425,517 | 20,414,307 |
| 19 | Christmas Food Court Flash Mob, Hallelujah Chorus | Alphabet Photography | 2010 | 2,394,191 | 42,525,502 |
| 20 | Thank You Mama - Best Job | P&G | 2012 | 2,287,408 | 12,629,369 |
| 21 | Telekinetic Coffee Shop Surprise | MGM | 2013 | 2,186,490 | 49,481,940 |
| 22 | Danny MacAskill - «Way Back Home» | Red Bull | 2010 | 2,114,377 | 29,260,802 |
| 23 | Ma Contrexpérience | Nestlé | 2011 | 2,041,967 | 19,735,431 |
| 24 | Write The Future | Nike | 2010 | 2,023,509 | 41,896,561 |
| 25 | Ape With AK-47 | 20th Century Fox | 2011 | 2,022,527 | 32,289,664 |

*The clip «9/11» by Budweiser was TV broadcast during the «Super Bowl» in 2002, but the video the data refers to appears in the Unruly Media database in 2008.
(Source: Unruly Media)

We thought that we should include the item «using celebrities», even if this was not mentioned in previous research, as this is a widely used resource in the history of advertising. We should clarify in this respect that we consider a «celebrity» any person who is well known in the concrete market or niche reflected or targeted by the video, and not necessarily at international level.

Through the quantification of the data related to the variables presented, and the cross-checks performed we have obtained the results that are presented below.

## 3. Analysis and outcomes

The first outcomes are of a descriptive nature, and allow us to present a basic profile of the average viral video ad within the group of the most successful ones. The conclusions drawn and relationships present across these elements will be presented in the following section:

a) Duration. Viral videos tend to last longer than most TV clips, with an average of around 03:05 minutes. However, we must highlight that in the sample taken, there are three videos of the same brand –DC Shoes– that increase the average significantly, as their durations are 09:52, 07:42 and 09:16. If we leave these particular clips aside, the average of the remaining sample ranges from 00:31 of the shortest video to 04:57 of the longest, and therefore the average in this second case is of 02:02.

b) Broadcast year. All viral videos in Unruly Media's Top 25 are above 2 Million shares. Despite the fact that we have taken the world ranking after 2006 as a reference, the first positions are taken by more recent videos against the oldest: of the 25 viral videos analysed. In the ranking between 2006 and 2013, 60% of the videos were aired in 2013 (9 videos) and 2012 (6 videos), while the others were aired in 2011 (4 videos), 2010 (4 videos), 2009 (1 video) and 2002-08 (1 video).

c) Target market. The data compiled show that 52% of viral videos analysed initially target an international market, however, 32% initially target (looking at the products or brands they advertise, or at the people in them, for example) the US market. The presence of viral video ads targeting other national markets is negligible, albeit present; of the four remaining videos (12%) «A Dramatic Surprise on a Quiet Square» addresses audiences in Belgium, the Netherlands and Luxembourg, «Dumb Ways to Die» targets an Australian audience, «Keyfi Yolunda, A?k? Sonunda», is produced for Cornetto Turkey, and «Ma Contrexpérience» targets the French market.

d) Sponsors. The most remarkable sponsors are DC Shoes, who manage to place their viral videos in 2012, 2010 and 2011 in the 25 most widely shared ones, Budweiser, who manage to have two of their «Super Bowl» videos (2002-13) in the ranking, and Evian, with two more clips (2009-13).

e) TV broadcasting. Regarding TV broadcasting of viral video ads, confirmation with absolute certainty of the broadcasting of some of the videos in the sample proved difficult. Therefore, we finally left this element out of our analysis. What we could confirm, however, was that three of them were broadcast during the «Super Bowl», amongst them the video ranking first.

f) Presence of humour, erotic or violent elements. While the presence of humorous content in the analysed videos is high (56%), erotic (8%) and violent (16%) content is low.

g) Ekman's basic emotions. According to our analysis, 76% of the most successful viral videos use surprise as a resource, and in all of them another basic emotion also appears; within this group, 37% correspond to the category «real stunt», 21% to «fictional stunt», 16% to «surprise event» and 26% show «narrative surprise». Regarding the remaining emotions, only happiness is used in a remarkable way as a resource in 92% of videos. We have not found any viral video ads in the sample using disgust or rage, while only 12% used fear to some extent (as a situation in the video) and 20%, sadness.

h) Presence of celebrities. The presence of celebrities is confirmed in 32% of the videos. In 87.5% of cases their presence is associated to the use of surprise as «real stunt».

i) Final emotional tone conveyed. The analysis of the final emotional tone conveyed through the combination of the elements present in the clips produces a result of «agreeable» in the total 25 viral video ads most widely shared in the period 2006-2013.

## 4. Discussion and conclusions

The most important contributions of our research have to do with the characterization of the most successful viral video ads –those that could be considered an example, at least from the point of view of the level of dissemination reached amongst users– and with the identification of some emotional and narrative elements in the videos, previously indicated by research on user perceptions of this content through quantitative methodologies.

Despite the fact that audiovisual content for the web is by definition supposed to have more formal and creative freedom than standard clips, the average duration of the most widely shared viral video ads is, as a whole, not far from the standard TV ad, if we leave the three longer pieces by DC Shoes aside, which could be considered exceptions to the sample, especially as they belong to the same sponsor (without them, the average duration would be of 02:02). However, the fact that the «Gymkhanas» of DC Shoes are so obviously present in the ranking leads us to think that if the content connects with the user, recommendations related to the duration of audiovisual pieces for Internet become secondary.

Research shows that 60% of the most widely shared viral video ads in the ranking from 2006-2013 were launched in the past two years. New entries in the Top 25 displace videos that show slow increases in the number of views and shares after they were launched. The impression this fact makes is that videos aired at present manage a high number of shares more easily than those produced in previous years, which means higher viral dissemination during launching. One probable cause for this is the unstoppable expansion of social networks, with more users by the day, and who enable the act of sharing with larger contact lists and therefore enable videos to gain greater coverage and impact. Another possibility –related to the first one– is that, once the maturity phase of the phenomenon of viral video ads was reached, companies started investing more actively in their dissemination, in an open or covert manner, optimising segmentation and «guiding virality». Both hypotheses leave the door open to specific research in this sense.

Another interesting result is that this viral phenomenon is not necessarily global in scope, at least in origin. It is true that certain brands launch their viral ads to an international market (52%), but it is also true that in many cases the large market share of a particular market, such as for example the US one, with its global media control, allows clips originally developed for this market (32%) to go international through events such as the «Super Bowl» or, simply, they just reach a high volume of shares without actually leaving their borders. A remarkable aspect is that only the Turkish ad «Keyfi Yolunda, A?k? Sonunda» uses a language other than English in its voice-over (a song). In fact, despite the fact that we did not take language as a specific feature in our analysis, we think that an interesting element would be to quantify the number of viral ads that remove verbal aspects to maybe favour internationalisation.

Despite the fact that there were several difficulties in compiling information that led us to do away with the item «TV broadcasting», we should highlight that the most widely shared video to date is an ad broadcast during the «Super Bowl»: «The Force», by Volkswagen. This in itself makes us question authors who argue that content of viral video ads is different and more «aggressive» than TV ads (Porter & Golan, 2002), or even the nature of virality because both «The Force», as well as viral videos by Budweiser, were initially broadcast on television during the «Super Bowl», and later posted on the web. This leads us to believe that television and events of mass audiences are a powerful starting point for the later viral dissemination of some of the most successful videos. Even though the initial concept of virality implies the use of networks of users in their dissemination, an interesting pending topic would be to re-examine whether this is still so.

The results obtained on the analysis of feelings are generally consistent with previous research based on user perception or behavioural prediction in the face of viral content.

Regarding the presence of humour, erotic (nudity) or violent elements in the videos analysed, the outcomes are different to some extent to those presented by Porter and Golan (2002). The presence of humorous content in the most widely shared viral adds is remarkable (58%), while erotic (8%) and violent (16%) content is present in reduced percentages, although it is true that whenever violence or eroticism are found, they are mainly presented in a humorous light, and only in one of the cases the use of erotic elements does not imply humour.

Regarding the analysis of Ekmar's basic emotions used by Dobele et al. (2007), the results obtained largely correspond with those presented by these researchers. 76% of the most widely shared viral videos use surprise as a resource, and all of them show at least another basic emotion.

Surprise through potentially dangerous activities related to sports or risky situations is a very common practice of the most successful viral video ads, and has much to do with the use of celebrities, as it is in these roles that such celebrities are usually cast if they are sportspeople or actors specialising in the practice they show in front of the camera. The outcomes of the analysis of surprise as an emotion in the most widely shared viral video ads coincide with the general approaches of previous research, such as those by Teixeira (2012) or Dobele et al. (2007), who find this the most relevant emotion in the success of viral videos.

We also consider that the division into four categories («real stunt», «fictional stunt», «surprise event» and «narrative surprise») proposed to classify surprise in viral video ads may be interesting both from a descriptive perspective and from the point of view of future research.

Regarding the remaining emotions analysed, only happiness appears in 92% of videos, as we have already explained. Regarding the emotions of fear and sadness, we have to highlight that fear is always used as a tool to generate surprise, so that the final feeling is positive. Regarding sadness, we also have to highlight that its use is mainly linked to happiness, generating an unexpected happy ending typical of videos using «narrative surprise» as a resource, and it is only in one case that we find sadness as a single emotion, in the clip «9/11», by Budweiser, of a clearly emotional nature.

Therefore, surprise and happiness are remarkable emotions in the sample of the most widely shared viral video ads, and this coincides with the contributions of Teixeira (2012), Dobele et al. (2007) or Eckler and Bolls (2011) in this regard.

On the other hand, and related to all of the above, the analysis of the final emotional tone conveyed by these ads shows that all viral videos in the Top 25 ranking can be classified as «agreeable», which coincides fully with the results of Eckler and Bolls (2011).

Our current research confronts the elements –provided by previous studies– seen as relevant by users when sharing content with the elements that are clearly present in the most widely shared viral video ads to date. The outcomes presented show a clear correlation between the findings of previous research from the point of view of users and the results obtained through content analysis of the 25 most successful cases of viral advertising in the past years, but they also open up questions that will allow us to make progress in future research.

## Notes

1 The concept of «media virus» by Rushkoff is related to the «meme», a unit of cultural transmission introduced by Dawkins in 1976, on whose evolution and interpretation Lull and Neiva (2011) have widely reflected.

2 On the origin of the approach of viral marketing and the different contributions to the coining of this term, please see Sivera (2008).